 \documentclass[final,3p,times,twocolumn]{elsarticle}
 
\usepackage{amssymb}

\usepackage{ulem}
\usepackage{color}

\newcommand{\beq}{\begin{eqnarray}}
\newcommand{\eeq}{\end{eqnarray}}
\newcommand{\be}{\begin{eqnarray*}}
\newcommand{\ee}{\end{eqnarray*}}

\newcommand{\ie}{{\it i.e.}}

\newcommand{\cf}[1]{{Fig.~\ref{#1}}}

\def\lsim{\raise0.3ex\hbox{$<$\kern-0.75em\raise-1.1ex\hbox{$\sim$}}}
\def\gsim{\raise0.3ex\hbox{$>$\kern-0.75em\raise-1.1ex\hbox{$\sim$}}}

\def\dAu  {$d$Au}

\def\pp   {$pp$}
\def\pA   {$pA$}

\def\AA   {$AA$}

\def\PbPbm {\mathrm{PbPb}}
\def\pPb  {$p$Pb}
\def\pPbm  {p\mathrm{Pb}}

\def\Npart   {\mbox{$N_{\rm part}$}}
\def\Ncoll   {\mbox{$N_{\rm coll}$}}

\def\RPbPb    {\mbox{$R_{\rm PbPb}$}}

\def\jpsi   {\mbox{$J/\psi$}}
\def\ccbar {\mbox{$c\bar{c}$}}
\def\pT      {\mbox{$P_{T}$}}
\def\kT     {\mbox{$k_{T}$}}
\def\sigabs {\mbox{$\sigma_{\mathrm{abs}}$}}
\def\beq     {\begin{equation}}
\def\eeq     {\end{equation}}
\def\upsi    {\mbox{$\Upsilon$}}

\begin{document}

\begin{frontmatter}

\title{Quarkonium as a tool: cold nuclear matter effects }


\author[address2]{E. G. Ferreiro}

\author[address3]{F. Fleuret}

\author[address4]{J.P Lansberg}

\author[address5]{N. Matagne}

\author[address6]{A. Rakotozafindrabe}

\address[address2]{Departamento de F\'{\i}sica de Part\'{\i}culas and IGFAE, Universidad de Santiago de Compostela, 
E-15782 Santiago de Compostela, Spain}
\address[address3]{LLR, \'Ecole polytechnique, CNRS, F-91128 Palaiseau, France}
\address[address4]{IPNO, Universit\'e Paris-Sud 11, CNRS/IN2P3, F-91406 Orsay, France}
\address[address5]{Universit\'e de Mons, Service de Physique Nucl\'eaire et Subnucl\'eaire, Place du Parc 20, B-7000 Mons, 
Belgium}
\address[address6]{IRFU/SPhN, CEA Saclay, F-91191 Gif-sur-Yvette Cedex, France}

\begin{abstract}
We discuss the quarkonium production as a tool for the study of the Quark Gluon Plasma. In particular, we concentrate on the Cold Nuclear Matter effetcs. We show that quarkonium production is also useful for the study of 
Quantum Chromodynamics first principles and the nuclear Parton Distribution Functions.
\end{abstract}

\begin{keyword}
Quarkonium  \sep Shadowing \sep Cold Nuclear Matter effects
\end{keyword}

\end{frontmatter}

\section{Introduction}
\label{Introduction}
The interest devoted to quarkonium production has not decreased for the last thirty years.
It is motivated by the search of the transition
from hadronic matter to a deconfined state of matter,
the so-called Quark-Gluon Plasma (QGP).
The high density of gluons in the QGP is expected to hinder the formation of quarkonium systems,
by a process analogous to Debye
screening of the electromagnetic field in a plasma \cite{Matsui86}.

Nevertheless, the BNL Relativistic Heavy Ion Collider data on $d$+Au collisions \cite{Adare:2007gn} have also
revealed that Cold Nuclear Matter (CNM) effects play an essential
role at these energies.

All this confirms that, in fact,
the interpretation of the results obtained in nucleus-nucleus collisions relies on a good
understanding and a proper subtraction of the CNM effects, known to
impact the quarkonium production in proton(deuteron)-nucleus collisions where the
deconfinement can not be reached.

We will focus on $d$+Au data, where no QGP formation is possible and 
only CNM effects are in play.
Two CNM effects have been identified
as important for quarkonium production: the nuclear absorption, commonly characterized
as
a final-state effect on the produced quarkonium state and parametrised through an effective 
breakup cross
section, and the shadowing, {\it i.e.} the modification of the
parton
densities in nuclei relative to the nucleon, an initial-state effect.

Moreover, 
we have recently realized that the theoretical uncertainty on the shadowing is not limited to the uncertainty 
on the modification of the parton distributions:
we do not know the specific production kinematics at a partonic level, 
this prevents us to know the exact momentum fractions of the gluon in the nucleus and thus at which $x_B$ to evaluate it.

In this work, we illustrate the importance of cold matter effects on $J/\psi$ production in $p$Pb and 
PbPb collisions at the
LHC and we recall that a proper subtraction of CNM effects on quarkonium production is mandatory before any QGP studies at
the LHC.


\section{Our approach}
\label{sec:approach}
As we have shown in earlier studies~\cite{Ferreiro:2008wc,Ferreiro:2009qr,Ferreiro:2009ur,Rakotozafindrabe:2010su}, considering the adequate \jpsi\ partonic production mechanism -- either via a \mbox{$2\to 1$} or a \mbox{$2\to 2$} process -- affects the way to compute the nuclear shadowing and thus its
expected impact on the \jpsi\ production.  In the following, we will refer to the first
scenario as the {\it intrinsic} scheme, and to the second as the {\it extrinsic} scheme.  Most studies on the \jpsi\ production in hadronic collisions are carried out in the intrinsic scheme. They rely on the assumption that
the \ccbar~pair is produced by the fusion of two gluons carrying
some intrinsic transverse momentum~\kT. The partonic process being a
\mbox{$2\to 1$} scattering, the sum of the gluon intrinsic \kT\ is transferred to the \ccbar~pair, thus to
the \jpsi\ since the soft hadronisation process does not alter
significantly the
kinematics.
 Thus, in such approaches, the transverse momentum \pT\ of the
\jpsi\ {\it entirely} comes from the intrinsic \kT\ of the initial gluons. However, the average value of~\kT\ is not expected to go much beyond
$\sim 1\mathrm{~GeV}$. So this process is not sufficient to describe the \pT\ spectrum of quarkonia in
hadron collisions~\cite{Lansberg:2006dh}.

For $\pT \, \gsim
\,2-3\mathrm{~GeV}$, most of the transverse momentum of the quarkonia should have an extrinsic
origin, \ie\ the \jpsi's \pT\ would be balanced by the emission of a recoiling particle in the final
state.
The \jpsi\ would then be produced by gluon fusion in a \mbox{$2\to 2$} partonic 
process with the emission of a hard final-state gluon.
This emission, which is anyhow mandatory to conserve $C$-parity,
has a definite influence on the kinematics of the
\jpsi\ production. Indeed, for a given \jpsi\ momentum (thus for
fixed rapidity~$y$ and \pT), the processes discussed above, \ie\ the intrinsic $g+g \to \ccbar \to J/\psi \,(+X)$
and the extrinsic $g+g \to J/\psi +g$,  will proceed on the average from initial gluons with different Bjorken-$x$. Therefore,
 they will be affected by different shadowing corrections.

In the intrinsic scheme, the measurement of the \jpsi\ momentum in \pp\ collisions completely fixes the longitudinal
 momentum fraction of the initial partons:
\begin{equation}
x_{1,2} = \frac{m_T}{\sqrt{s_{NN}}} \exp{(\pm y)} \equiv x_{1,2}^0(y,P_T),
\label{eq:intr-x1-x2-expr}
\end{equation}
with $m_T=\sqrt{M^2+P_T^2}$, $M$ being the \jpsi\ mass.

\begin{figure}[htb!]
\begin{center}
\includegraphics[width=7cm]{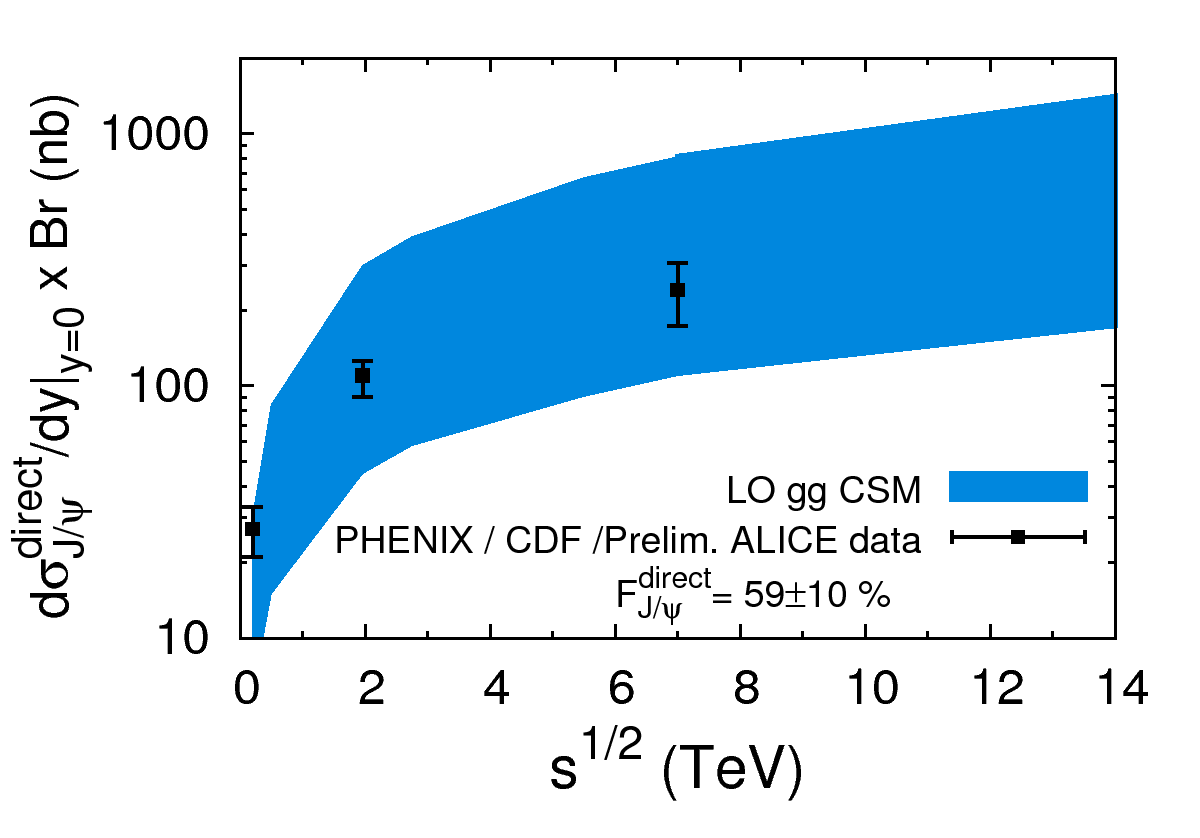}
\end{center}
\vspace{-0.8cm}
\caption{$d\sigma^{\mathrm{direct}}_{J/\psi}/dy|_{y=0}\ \times$ Br from $gg$ fusion in $pp$ collisions for 
$\sqrt{s}$ from 200 GeV up to 14 TeV compared to the PHENIX \cite{Adare:2006kf}, CDF \cite{Acosta:2004yw} and ALICE data.}
\label{fig:plot-dsigvsdy-y_0-231010}
\end{figure}

However, in the extrinsic scheme, the knowledge of
the $y$ and \pT\ spectra is not sufficient to determine $x_1$ and $x_2$.
Actually, the presence of a final-state gluon introduces further degrees
of freedom in the kinematics, allowing several $(x_1, x_2)$ for a given set $(y, P_T)$.
 The four-momentum conservation explicitely results in a more complex expression of $x_2$ as a function of~$(x_1,y,P_T)$:
\begin{equation}
x_2 = \frac{ x_1 m_T \sqrt{s_{NN}} e^{-y}-M^2 }
{ \sqrt{s_{NN}} ( \sqrt{s_{NN}}x_1 - m_T e^{y})} \ .
\label{eq:x2-extrinsic}
\end{equation}
Equivalently, a similar expression can be written for $x_1$ as a function of~$(x_2,y,P_T)$.
Even if the kinematics determines
the physical phase space, models are anyhow {\it mandatory} to compute the proper
weighting of each kinematically allowed $(x_1, x_2)$. This weight is simply
the differential cross section at the partonic level times the gluon PDFs,
\ie\ $g(x_1,\mu_F) g(x_2, \mu_F) \, d\sigma_{gg\to J/\psi + g} /dy \,
dP_T\, dx_1 dx_2 $.
In the present implementation of our code, we are able to use the partonic differential
cross section computed from {\it any} theoretical approach. In this work, we shall use the Colour-Singlet Model (CSM) at LO at LHC energy, which was shown to be compatible~\cite{Brodsky:2009cf,Lansberg:2010cn} 
--see Fig. \ref{fig:plot-dsigvsdy-y_0-231010} for illustration-- 
with the magnitude of the \pT-integrated cross-section as given by the PHENIX \pp\ data~\cite{Adare:2006kf}, the CDF $p\bar{p}$ data~\cite{Acosta:2004yw} and the recent LHC \pp\ data at $\sqrt{s_{NN}} = 7\mathrm{~TeV}$.

To obtain the yield of $J/\psi$ and $\Upsilon$ in \pA\ and \AA\ collisions, a shadowing-correction
factor has to be applied to the \jpsi\ yield obtained from the simple
superposition of the equivalent number of \pp\ collisions.
This shadowing factor can be expressed in terms of the ratios $R_i^A$ of the
nuclear Parton Distribution Functions (nPDF) in a nucleon belonging to a nucleus~$A$ to the
PDF in the free nucleon:

\begin{equation}
\label{eq:shadow-corr-factor}
R^A_i (x,Q^2) = \frac{f^A_i (x,Q^2)}{ A f^{nucleon}_i (x,Q^2)}\ , \ \
i = q, \bar{q}, g \ .
\end{equation}
The numerical parameterisation of $R_i^A(x,Q^2)$
is given for all parton flavours. Here, we restrict our study to gluons since, at
high energy, the quarkonia are essentially produced through gluon fusion
\cite{Lansberg:2006dh}. Several shadowing parametrisations are available~\cite{deFlorian:2003qf,Eskola:1998df,Eskola:2008ca,Eskola:2009uj}. 
In the following, we shall restrict ourselves to EKS98~\cite{Eskola:1998df}, which is very close to the mean in the current evaluation of the uncertainty~\cite{Eskola:2009uj} on the gluon nPDF and exhibit a moderate antishadowing. 

The second CNM effect that we take into account concerns
the nuclear absorption.  In the framework of the probabilistic Glauber
model, this effect is usually parametrised
by introducing an effective absorption cross
section~\sigabs. It reflects the break-up of correlated \ccbar~pairs due to inelastic scattering with the remaining nucleons from the incident cold nucleus. The value of~\sigabs\ is unkown at the 
LHC. At high energy, the heavy state in the projectile should undergo a coherent scattering off the nucleons of the target nucleus~\cite{Kopeliovich:2001ee}, in contrast to the incoherent, longitudinally ordered scattering that takes place at low energies. As argued in~\cite{Capella:2006mb,Tywoniuk:2007gy}, this should lead to a decrease of \sigabs\ with increasing $\sqrt{s_{NN}}$. Compilation and systematic study of many experimental data indicate that \sigabs\ appears either constant~\cite{Arleo:2006qk} or decreasing~\cite{Lourenco:2008sk} with energy. Hence, we can consider our estimates~\cite{Ferreiro:2009ur,Rakotozafindrabe:2010su} of \sigabs\ at RHIC energy as upper bounds for the value of \sigabs\ at the LHC. 
We choose three values of the absorption cross section that should span this 
interval ($\sigma_{\mathrm{abs}} = 0, 1.5, 2.8\mathrm{~mb}$).

\section{Results}
\label{sec:results}
In the following, we present our results for the \jpsi\ nuclear modification factor due to CNM effects in the extrinsic sheme in \pPb\ and PbPb collisions at $\sqrt{s_{NN}} = 5.5\mathrm{~TeV}$: $R_{AB} = dN_{AB}^{J/\psi}/\langle\Ncoll\rangle dN_{pp}^{J/\psi}$, where $dN_{AB}^{J/\psi} (dN_{pp}^{J/\psi})$ is the observed \jpsi\ yield in $AB = \pPbm,\PbPbm$ (\pp) collisions and $\langle\Ncoll\rangle$ is the average number of nucleon-nucleon collisions occurring
in one \pPb\ or PbPb collision. Without nuclear effects, $R_{AB}$ should equal unity.
\begin{figure}[htb!]
\begin{center}
\includegraphics[width=6cm]{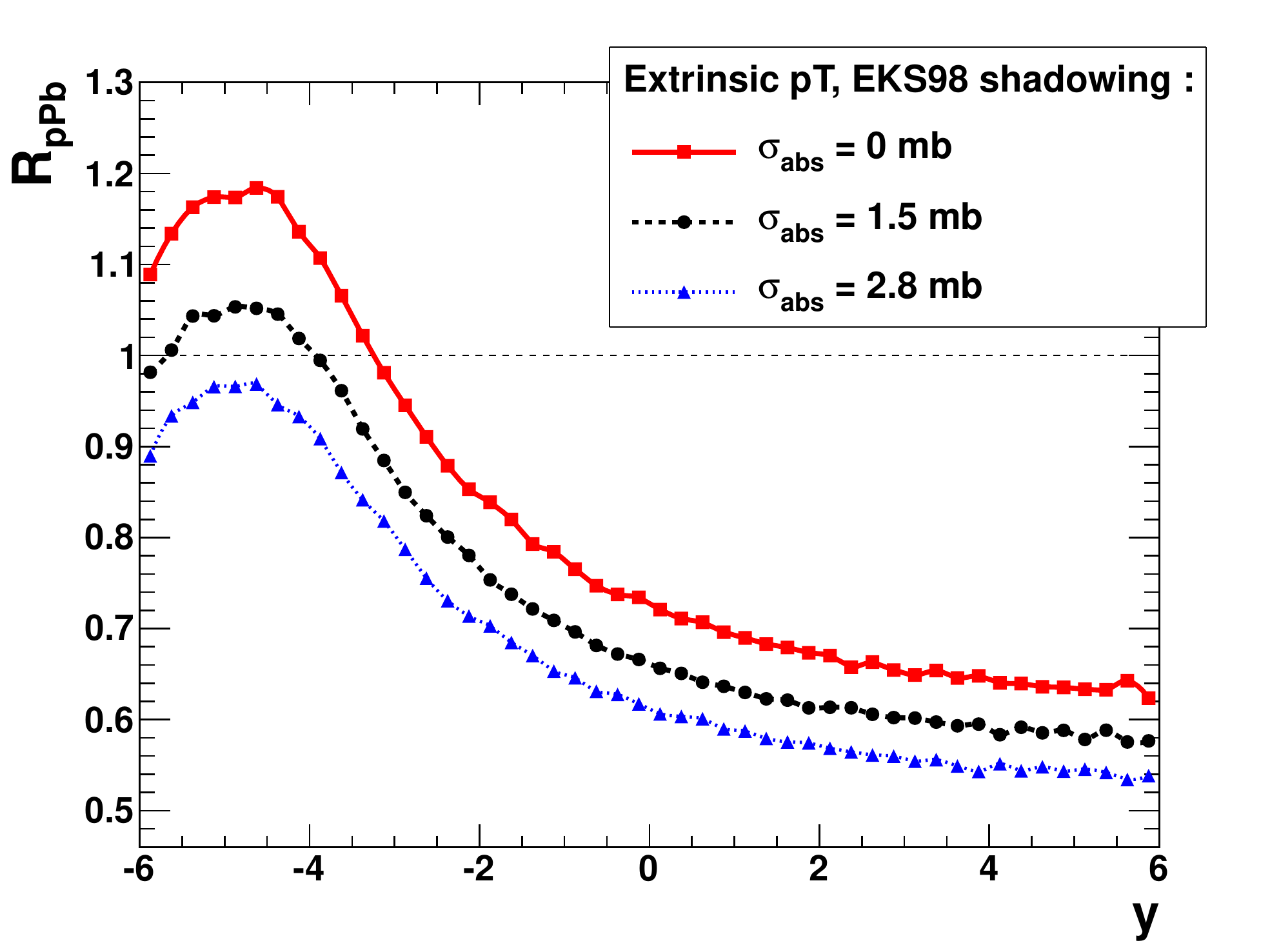}
\includegraphics[width=6cm]{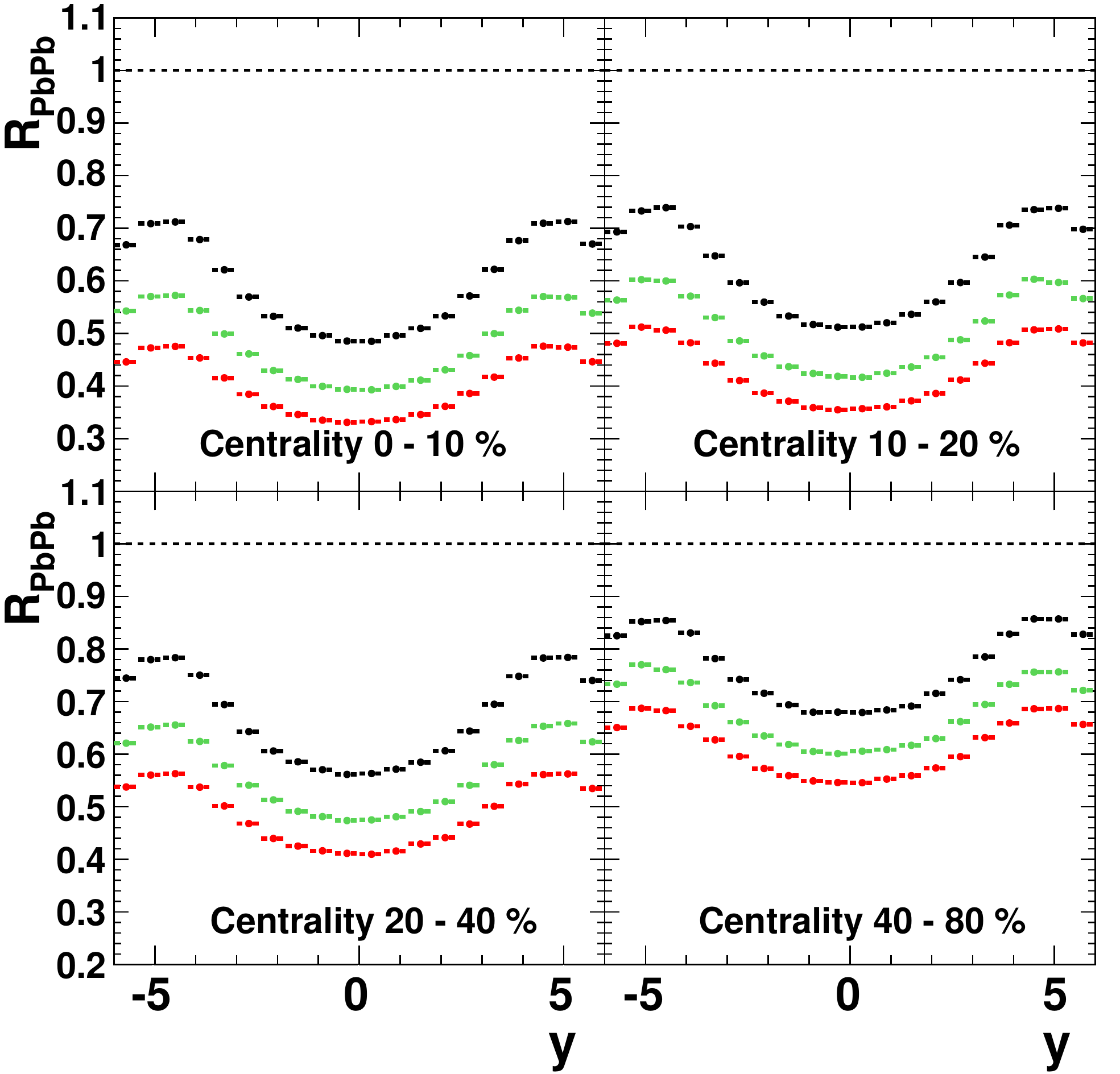}
\end{center}
\vspace{-0.6cm}
\caption{\jpsi\ nuclear modification factor versus $y$ in \pPb\ and PbPb collisions at $\sqrt{s_{NN}}=5.5\mathrm{~TeV}$, using EKS98~\cite{Eskola:1998df} gluon shadowing parametrisation and three values of $\sigma_{abs}$ (from top to bottom: $0, 1.5, 2.8\mathrm{~mb}$) in the extrinsic scheme. For PbPb collisions, the $y$-dependence is shown for various centrality selections.}
\label{fig:RpPb_vs_y_and_RPbPb_vs_y_vs_cent}
\end{figure}

\begin{figure}[htb!]
\begin{center}
\includegraphics[width=6cm]{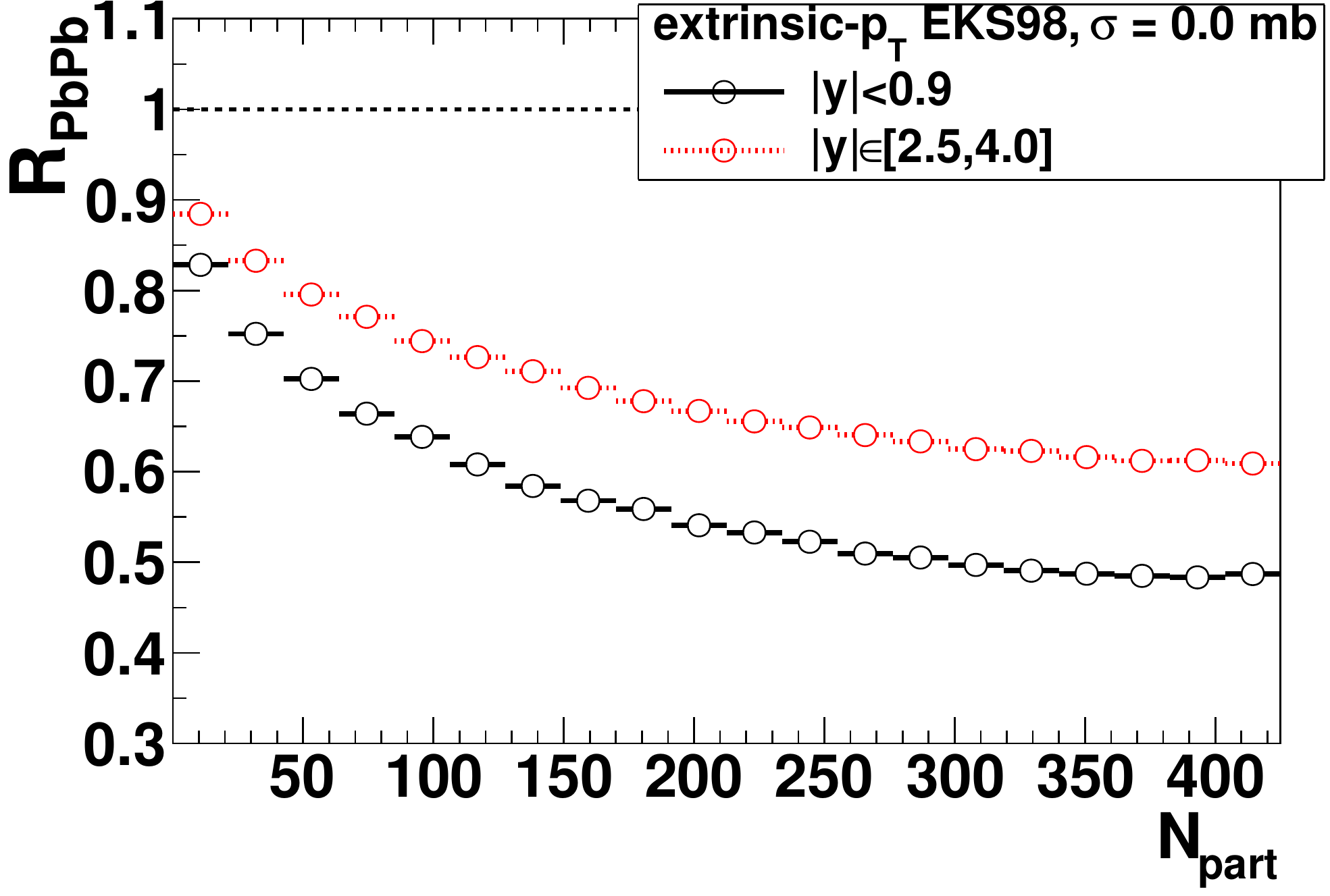}
\includegraphics[width=6cm]{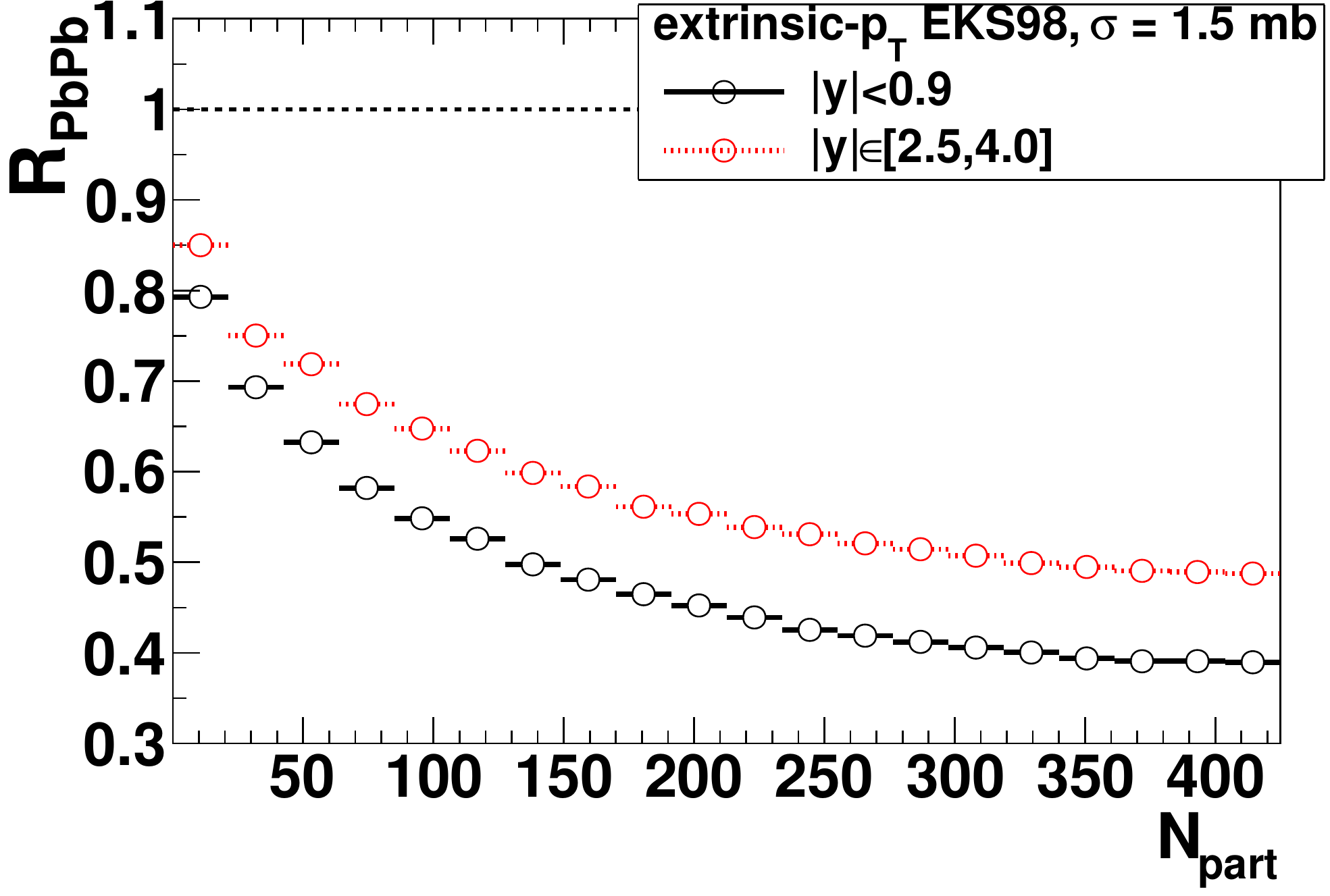}
\includegraphics[width=6cm]{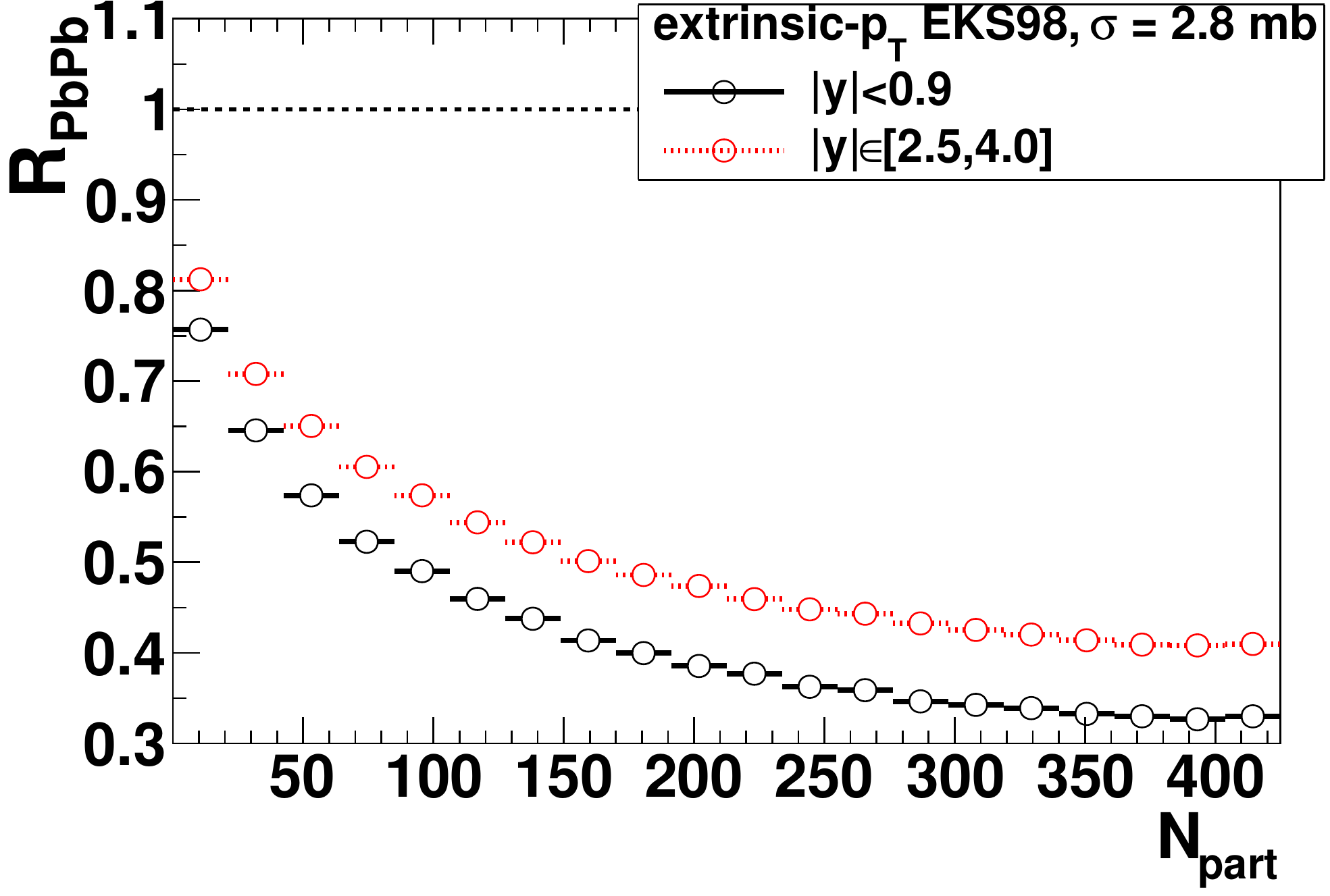}
\end{center}
\vspace{-0.6cm}
\caption{(Color online) \jpsi\ nuclear modification factor, \RPbPb, in PbPb collisions at $\sqrt{s_{NN}}=5.5\mathrm{~TeV}$ versus \Npart, using EKS98~\cite{Eskola:1998df} gluon shadowing parametrisation and three values of the nuclear absorption cross section in the extrinsic scheme. \RPbPb\ is shown for two different experimental acceptances in rapidity, $|y|<0.9$ and $|y|\in [2.5, 4]$.}
\label{fig:RPbPb_vs_Npart}
\end{figure}
In \cf{fig:RpPb_vs_y_and_RPbPb_vs_y_vs_cent}, we show $R_{\pPbm}$ versus $y$. The curve with no absorption allows us to highlight the strong rapidity dependence of the shadowing. We can also notice that the shadowing alone should already be responsible for a quite large amount of \jpsi\ suppression, up to \mbox{$36\,\%$} at $y=6$. 
This is expected because one accesses at LHC energy the region of very small x in the gluon nPDF
(down to $10^{-5}$). At backward rapidity, we are in the antishadowing region, with the antishadowing peak at $y \simeq -5$. \cf{fig:RpPb_vs_y_and_RPbPb_vs_y_vs_cent} shows that the $y$-dependence of $R_{\PbPbm}$ for the centrality bins used by the ATLAS Collaboration \cite{ATLAS:2010px}
is similar for all the bins, with a dip at mid-$y$. This shape is the opposite of the one obtained at RHIC energy~\cite{Ferreiro:2008wc,Ferreiro:2009ur}, with a peak at mid-$y$.
One can also notice that $R_{\PbPbm}$ decreases at very large rapidity showing that one has gone beyond the antishadowing peak. Here, $R_{\PbPbm}$ is systematically smaller at mid-$y$ than at forward-$y$. This is also illustrated on \cf{fig:RPbPb_vs_Npart}, with the centrality dependence of $R_{\PbPbm}$ for two regions in~$y$.
This behaviour of the CNM effects may partially -- or completely -- compensate the opposite effect expected from \ccbar\ recombination, with a maximum enhancement at $y=0$. We should however note that the recombination of $c\bar c$ pairs is expected to be less important if an experimental  $P_T$ cut has to be imposed because of a limited acceptance at low $P_T$.
Overall, one may observe a $R_{\PbPbm}$ rather independent of $y$ resulting of two $y$-dependent effects.

\begin{figure}[htb!]
\begin{center}
\vspace{-0.2cm}
{\includegraphics[width=6cm]{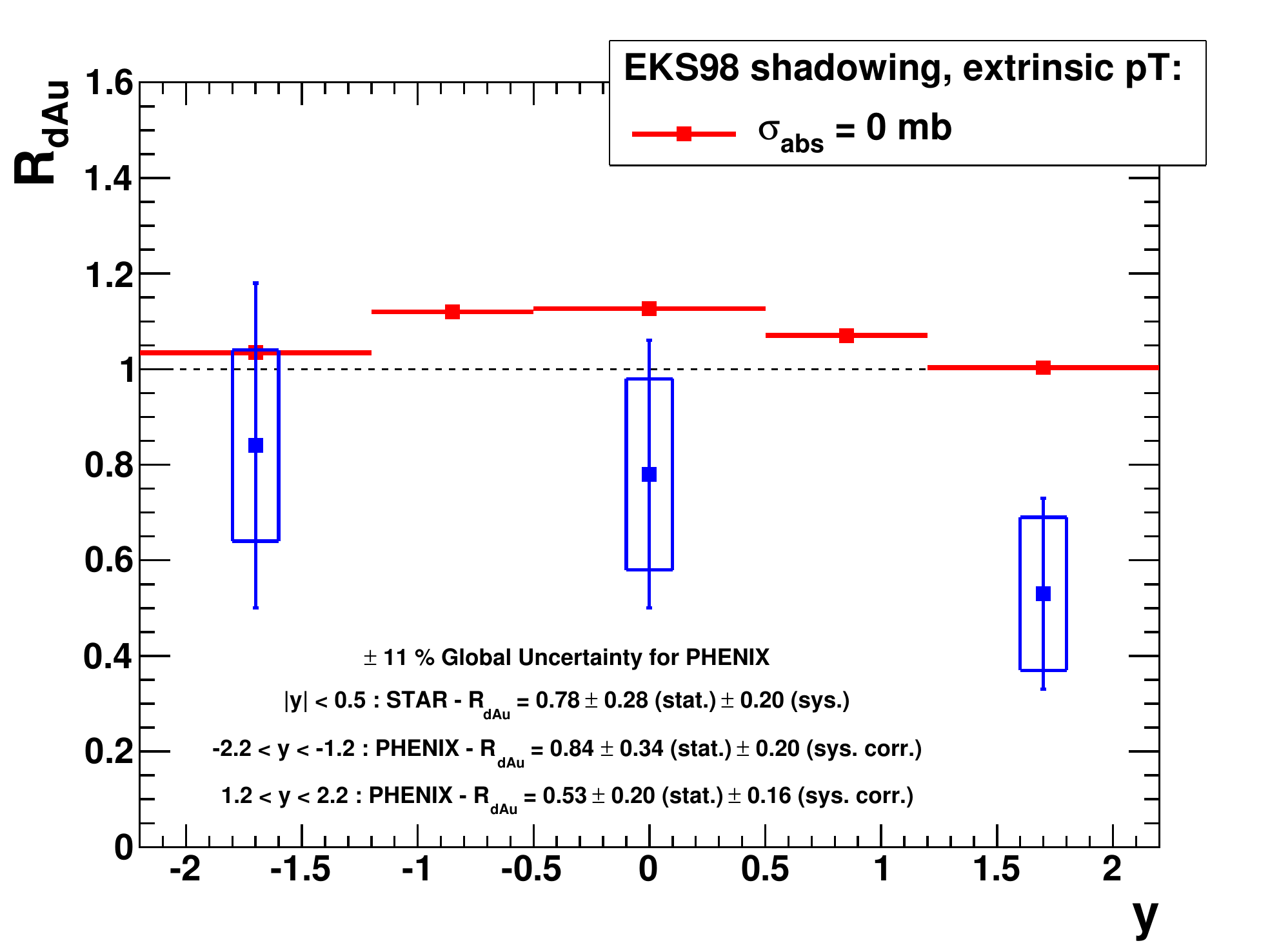}}
{\includegraphics[width=6cm]{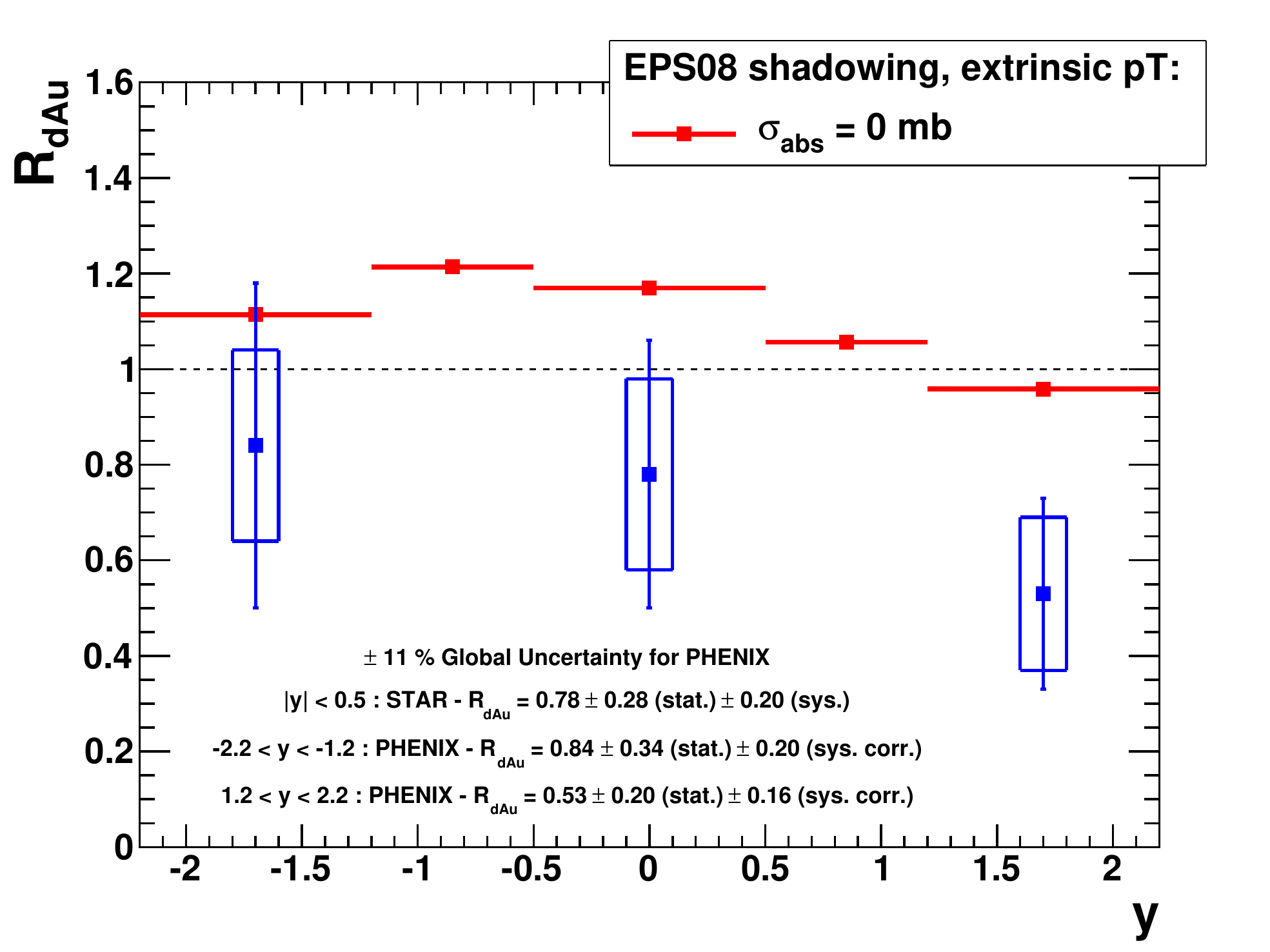}}
{\includegraphics[width=6cm]{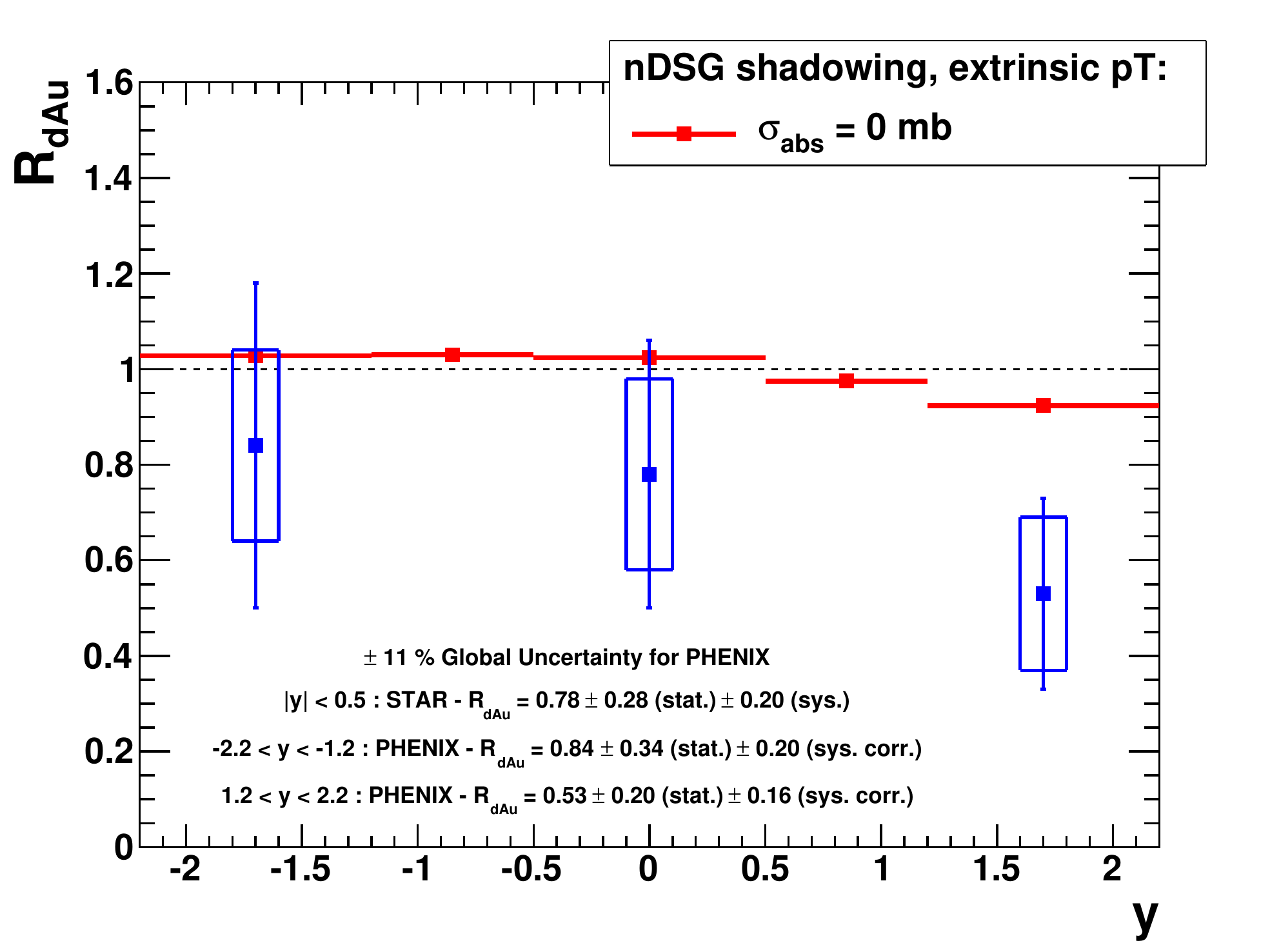}}
\end{center}
\vspace{-0.6cm}
\caption{\upsi\ nuclear modification factor in \dAu\ collisions at
$\sqrt{s_{NN}}=200\mathrm{~GeV}$ versus $y$
 using
the following gluon shadowing parametrisations: a) EKS98, b) EPS08, c) nDSg}
\label{fig:RdAu_vs_yUps}
\end{figure}

Recently, we have also paid attention to the preliminary data on \upsi\ production at RHIC. 
In Fig.~\ref{fig:RdAu_vs_yUps}, we show $R_{dAu}$ vs $y$ according to the
extrinsic scheme. The results are displayed
for 3 values of $\sigma_{\mathrm{abs}}$ and for three shadowing parametrisations and
these are compared to the available RHIC data~\cite{phenix,Liu:2009wa}.

\section{Conclusions}
\label{sec:conclusions}
\vspace{-0.2cm}
In conclusion, we know from lattice QCD that strongly interacting matter undergoes
a deconfinement transition to a new state, the QGP.
The objective of high energy nuclear collisions is to produce and study
the QGP under controlled conditions in the laboratory. 
The study of quarkonium production and
suppression is among the most interesting investigations in this field since
calculations indicate that the QCD binding potential is screened in the QGP
phase.
As demonstrated in this work, other phenomena will also
impact on the quarkonium production rates in nuclear collisions at the LHC.
For instance, nuclear parton distributions modifications and CNM
absorption of quarkonium  must be accounted for before any QGP studies at the LHC become
meaningful.
%
E.G.F. thanks Xunta de Galicia
and Ministerio de Ciencia of Spain
(FPA2008-03961-E/IN2P3), N.M. thanks the F.R.S.-FNRS (Belgium). 

\vspace{-0.1cm}
\bibliographystyle{elsarticle-num}

\vspace{-0.1cm}
\begin{thebibliography}{00}

\vspace{-0.1cm}
\bibitem{Matsui86}
  T.~Matsui and H.~Satz,
  Phys.\ Lett.\  B {\bf 178} (1986) 416.

\bibitem{Adare:2007gn}
  A.~Adare {\it et al.}, 
  Phys.\ Rev.\  C {\bf 77} (2008) 024912 

\bibitem{Ferreiro:2008wc}
   E.~G.~Ferreiro, F.~Fleuret, J.~P.~Lansberg and A.~Rakotozafindrabe,
   Phys.\ Lett.\ B {\bf 680} (2009) 50.

\bibitem{Ferreiro:2009qr}
   E.~G.~Ferreiro, F.~Fleuret, J.~P.~Lansberg and A.~Rakotozafindrabe,
   {\it 2009 QCD and High Energy Hadronic Interaction}, The gioi Publishers (2009) p.\ 395-398,
   arXiv:0903.4908 [hep-ph].

\bibitem{Ferreiro:2009ur}
  E.~G.~Ferreiro, F.~Fleuret, J.~P.~Lansberg and A.~Rakotozafindrabe,
  Phys.\ Rev.\  C {\bf 81} (2010) 064911.

\bibitem{Rakotozafindrabe:2010su}
  A.~Rakotozafindrabe, E.~G.~Ferreiro, F.~Fleuret and J.~P.~Lansberg,
  J.\ Phys.\ G {\bf 37} (2010) 094055.

\bibitem{Lansberg:2006dh}
  J.~P.~Lansberg, Int.\ J.\ Mod.\ Phys.\ A {\bf 21} (2006) 3857.

\bibitem{Brodsky:2009cf}
  S.~J.~Brodsky and J.~P.~Lansberg,
  Phys.\ Rev.\  D {\bf 81} (2010) 051502.

\bibitem{Lansberg:2010cn}
  J.~P.~Lansberg,
  PoS {\bf ICHEP 2010}, 206 (2010)
  [arXiv:1012.2815 [hep-ph]].

\bibitem{Adare:2006kf}
  A.~Adare {\it et al.}, 
  Phys.\ Rev.\ Lett.\  {\bf 98} (2007)  232002.

\bibitem{Acosta:2004yw}
  D.~Acosta {\it et al.}  [CDF Collaboration],
  Phys.\ Rev.\  D {\bf 71} (2005) 032001.

\bibitem{deFlorian:2003qf}
  D.~de Florian and R.~Sassot,
  Phys.\ Rev.\  D {\bf 69} (2004) 074028.

\bibitem{Eskola:1998df}
  K.~J.~Eskola, V.~J.~Kolhinen and C.~A.~Salgado,
  Eur.\ Phys.\ J.\  C {\bf 9} (1999) 61.

\bibitem{Eskola:2008ca}
  K.~J.~Eskola, H.~Paukkunen and C.~A.~Salgado,
  JHEP {\bf 0807} (2008) 102.


\bibitem{Eskola:2009uj}
  K.~J.~Eskola, H.~Paukkunen and C.~A.~Salgado,
  JHEP {\bf 0904} (2009) 065.

\bibitem{Kopeliovich:2001ee}
  B.~Kopeliovich, A.~Tarasov and J.~Hufner,
  Nucl.\ Phys.\  A {\bf 696} (2001) 669.

\bibitem{Capella:2006mb}
  A.~Capella, E.~G.~Ferreiro, A.~Capella and E.~G.~Ferreiro,
  Phys.\ Rev.\  C {\bf 76} (2007) 064906.

\bibitem{Tywoniuk:2007gy}
  K.~Tywoniuk, I.~Arsene, L.~V.~Bravina, A.~B.~Kaidalov and E.~E.~Zabrodin,
  J.\ Phys.\ G {\bf 35} (2008) 044039.

\bibitem{Arleo:2006qk}
  F.~Arleo and V.~N.~Tram,
  Eur.\ Phys.\ J.\  C {\bf 55} (2008) 449.

\bibitem{Lourenco:2008sk}
  C.~Lourenco, R.~Vogt and H.~K.~Woehri,
  JHEP {\bf 0902} (2009) 014.

\bibitem{ATLAS:2010px}
  ATLAS Collaboration,
  CERN-PH-EP-2010-090, arXiv:1012.5419 [hep-ex].

\bibitem{phenix}
L. A. Linden Levy for PHENIX collaboration,
International Conference on High Energy Physics (ICHEP), Paris, France, July 21-28, 2010.

\bibitem{Liu:2009wa}
  H.~Liu  [STAR Collaboration],
  Nucl.\ Phys.\  A {\bf 830} (2009) 235C
  [arXiv:0907.4538 [nucl-ex]].

 \end{thebibliography}



\end{document}